\documentclass[journal]{IEEEtran}
\ifCLASSINFOpdf
\usepackage[pdftex]{graphicx}

\else

\fi

\usepackage[cmex10]{amsmath}
\usepackage{amssymb}   
\usepackage{amsxtra}
\usepackage{amscd}
\usepackage{amsthm}
 \usepackage{amsxtra}
 \usepackage{amscd}
 \usepackage{amsthm}
\usepackage{tikz}
\usetikzlibrary{matrix}
\usepackage{graphicx, subfigure}   
\usepackage{comment}
\usepackage{verbatim} 

\usepackage{amsmath}

\usepackage{array}  
\usepackage{authblk}
\usepackage{multirow}  
\usepackage{arydshln}

\usepackage{color}

\usepackage{cite}
\renewcommand{\thefootnote}{\arabic{footnote}}
\usepackage{balance}
\usepackage[keeplastbox]{flushend}

  \author{Emre Arslan,~\IEEEmembership{Student Member,~IEEE,} Ali Tugberk Dogukan,~\IEEEmembership{Student Member,~IEEE,} and \\Ertugrul Basar,~\IEEEmembership{Senior Member,~IEEE} 
\thanks{The authors are with the Communications Research and Innovation Laboratory (CoreLab),  Department of Electrical and Electronics Engineering, Ko\c{c} University, Sariyer 34450, Istanbul, Turkey. \mbox{Email: ebasar@ku.edu.tr, earslan18@ku.edu.tr, adogukan18@ku.edu.tr}} 
}\vspace{-0.5cm}

\begin{document}
	
	\title{ Sparse-Encoded Codebook Index Modulation  
	}

	\maketitle	
\vspace{-0.3cm}

	\begin{abstract}
	
Ultra-reliable and low-latency communications (URLLC) partakes a major role in 5G networks for mission-critical applications. Sparse vector coding (SVC) appears as a strong candidate for future URLLC networks by enabling superior performance in terms of bit error rate (BER). SVC exploits the virtual digital domain (VDD) and compressed sensing (CS) algorithms to encode and decode its information through active symbol indices. In this paper, first, a clever encoding/decoding algorithm is proposed for the SVC scheme, which allows the use of all possible activation patterns (APs) resulting in increasing spectral efficiency. Second, a novel solution is proposed to convey additional information bits by further exploiting index modulation (IM) for the codebooks of the SVC scheme. Computer simulation results reveal that our low-complexity algorithm and novel IM solution provide not only a superior BER performance but also an increase in the number of bits conveyed by IM compared to the ordinary SVC approach.





	\end{abstract}

	\IEEEpeerreviewmaketitle
		\IEEEpubidadjcol
		
	\renewcommand{\thefootnote}{\fnsymbol{footnote}}
	\section{Introduction}
\IEEEPARstart{O}{rthogonal} frequency division multiplexing (OFDM) is the most popular multi-carrier transmission technique due to its simplicity in generation and recovery, high spectral efficiency, multiple-input multiple-output (MIMO) compatibility and robustness to inter-symbol interference. Hence, OFDM is the currently used waveform in 4G, while 5G utilizes a flexible multi-numerology OFDM scheme. Great deal research has been done on OFDM and some have been very successful such as OFDM with index modulation (OFDM-IM), where subcarriers are grouped and data is transmitted through the indices of the active subcarriers along with $M$-ary signal constellations as in classical OFDM \cite{bacsar2013orthogonal,jaradat2018ofdm,bacsar2015ofdm}. However, even OFDM-IM alone, does not meet ultra-reliable and low-latency communication (URLLC) requirements since it does not provide diversity in the frequency domain.  
	
URLLC is a critical service for 5G applications. Future applications such as autonomous vehicles, wireless robotic surgery and real-time virtual reality entertainment demand extremely low block error rate (BLER) performance, as low as $10^{-5}$, with minimal latency on the order of milliseconds or less. As an example, in contrast to traditional manned vehicles, autonomous vehicles are extremely dependent on URLLC and waveform designs in 5G vehicular
networks \cite{URRLLCveh,vehref}. Numerous techniques and algorithms are being developed to meet these stringent requirements. However, it is challenging to have a highly reliable and low-latency system concurrently because of the trade-offs among them. Channel coding and various diversity techniques are widely used today to achieve a higher bit error rate (BER) performance at the expense of lower spectral efficiency and increased complexity. In addition, re-transmission is used for increasing reliability while introducing further latency.
	 
Recently, a number of methods have been implemented for OFDM to increase its reliability, such as compressed sensing (CS) and sparse vector coding (SVC) as given in \cite{zhang2016compressed} and \cite{ji2018sparse}, respectively. There have been many interesting CS-based studies dealing with sparse matrices, successful recovery with a high probability and sparse graph codes \cite{r11,r12,r13}. SVC appears as an attractive tool for OFDM by enabling outstanding BLER values due to its extremely sparse signal structure and CS-based ultra-reliable detection. Studies include various iterative sparse approximation algorithms such as the greedy ones, orthogonal matching pursuit (OMP) and multipath matching pursuit with depth\textcolor{red}{-}first (MMP-DF) \cite{kwon2014multipath}. These algorithms provide robust sparsity recovery in low signal-to-noise ratio (SNR) regions compared to previous approaches. Nevertheless, state-of-the-art schemes do not provide low-latency solutions and exhibit low data rates when using SVC with OFDM, which should be critical for 5G and beyond URLLC applications.

This paper aims to improve the latency, spectral efficiency and BER of plain SVC-based OFDM schemes for URLLC applications. Against this background, the contributions of this paper are twofold. First, a clever encoding/decoding technique is proposed for SVC-OFDM, which is referred to as \textit{enhanced SVC-OFDM (E-SVC-OFDM)}. E-SVC-OFDM increases the number of bits transmitted by IM in an intelligent manner. The MMP-DF algorithm is used for detection and since the sparsity of the system is high, it can be considered as a system with low complexity as in \cite{kwon2014multipath}. SVC-OFDM exploits a certain number of possible activation patterns (APs) of the active indices and provides multiple sparse vectors to use for data transmission. However, it is not always possible to use all possible APs since the total number of combinations is not an integer power of two. To solve this problem, E-SVC-OFDM runs a clever algorithm that allows the use of all possible APs, hence, the number of bits transmitted by APs is also increased. Second, a novel scheme called \textit{sparse-encoded codebook IM (SE-CBIM)} is introduced to further increase the number of bits transmitted by IM. In this scheme, in addition to APs, IM is used also for codebook indices with clever encoding and low-complexity decoding algorithms. 
	
\begin{figure*}[t]
\centering
\includegraphics[width=0.75\linewidth,height=3.0cm]{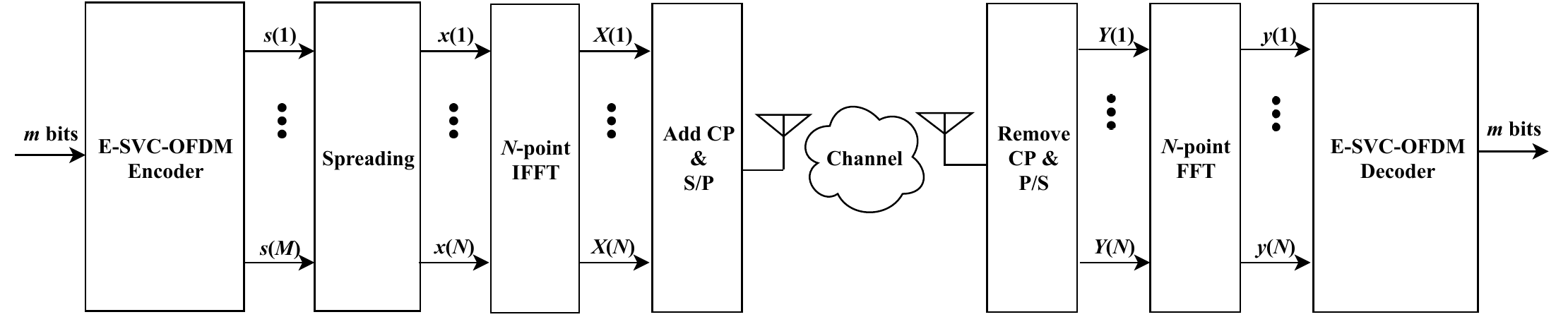}
\caption{System model of E-SVC-OFDM}
\vspace{-0.3cm}
\label{Fig. 1}
\end{figure*}
	
The rest of the paper can be summarized as follows. In Section II, the system model of E-SVC-OFDM is presented. In Section III, the SE-CBIM technique is proposed. Additionally, Section IV provides a complexity analysis. In Section V, computer simulation results are presented to observe the effects of varying parameters and to verify the superior performance of the proposed schemes. The paper is concluded in Section VI.


\section{Enhanced Sparse Vector Coding OFDM}

\subsection{E-SVC-OFDM Encoder}
The block diagram of the considered E-SVC-OFDM system is shown in Fig. 1. Firstly, using the combinatorial method \cite{bacsar2013orthogonal}, a sparse vector $\mathbf{s}\in\mathbb{C}^{M\times 1}$, which has $K$ non-zero elements with indices $(i_1,i_2,\cdots,i_K)$, where $i_k \in \{1,\cdots,M\}$ is the $k$th index, is constructed in the virtual digital domain (VDD)\cite{zhang2016compressed} according to the decimal equivalent ($d$) of the incoming $m$ bits via binary to decimal conversion. Once $(i_1,i_2,\cdots,i_K)$ are selected, $K$ constant unit-energy symbols that do not convey information, are assigned to these active indices similar to space shift keying (SSK) schemes \cite{article}, where the vector of these symbols is $\mathbf{b}_1=\begin{bmatrix}  x_1 & x_2 & \cdots & x_K \end{bmatrix}^\mathrm{T}$ and $[.]^\mathrm{T}$ denotes the transpose. A total of $ \binom{M}{K}$ APs, which can be generated by the combinatorial method and takes $d$ as an input, are available, where  $ \binom{\cdot}{\cdot}$ is the binomial coefficient. However, only $2^{ \lfloor a\rfloor }$ APs can be used, which results in a waste of potential use of all APs, where $a=\log_2\binom{M}{K}$ and ${ \lfloor .\rfloor }$ is the floor function. At this point, a clever technique that reuses the remaining $2^{ \lfloor a\rfloor+1}-\binom{M}{K}$ APs by extending the set of constant symbols as
$\mathbf{b}_2=\begin{bmatrix} \Tilde{x}_1 & \Tilde{x}_2 & \cdots & \Tilde{x}_K \end{bmatrix} ^\mathrm{T}$, is proposed. Therefore, a total of $m={ \lfloor a\rfloor}+1$
bits can be transmitted for each OFDM symbol. This technique enables us to use all possible APs and to transmit one additional bit. If $d$ is greater than $\binom{M}{K}-1$, the indices corresponding to $d-\binom{M}{K}$ are reused by extending the constant symbol set. We also avoid illegal APs in this way.

\begin{figure*}[t]
\centering
\includegraphics[width=1.00\linewidth,height=3.0cm]{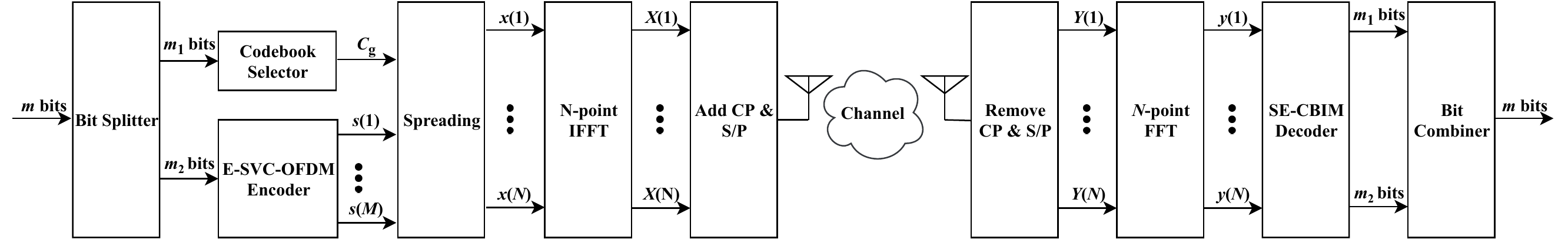}	
\caption{System model of SE-CBIM}
\vspace{-0.3cm}
\label{Fig. 1}
\end{figure*}
 
 As an example, assume two constant and unit-energy symbols $x_1=1$ and $x_2=j$ are transmitted via active indices for $M=4$ and $K=2$. All possible APs can be exploited with the reuse of the first two APs by extending the set of the constant symbols by $\Tilde{x}_1=-1$ and $\Tilde{x}_2=-j$ as shown in Table I. In the absence of this extension, we have a total of $6$ APs and only $2^{\lfloor\log_2{6}\rfloor}=4$ out of them can be employed to convey information. Here, the remaining $2$ APs are illegal. As seen from Table I, our proposed method enables the use of all possible APs and the transmission of $\log_2{8}=3$ bits.

 \begin{table}[t]
     \renewcommand{\arraystretch}{1.1}%
     \centering
     \caption{Index combinations according to incoming bits}
     \begin{tabular}{|>{\centering\arraybackslash}m{1cm} | >{\centering\arraybackslash}m{1cm}|>{\centering\arraybackslash}m{1cm} |>{\centering\arraybackslash}m{1cm} |>{\centering\arraybackslash}m{1cm} |>{\centering\arraybackslash}m{1cm} |} 
      	
  	\hline
    Bits & $d$ & First Index $(i_1)$ & Second Index $(i_2)$ & First Symbol & Second Symbol \\ [0.5ex]
      		
      \hline\hline
    $[000]$ & $0$ & $1$ & $2$ & $1$ & $j$ \\\hline
    $[001]$ & $1$ & $1$ & $3$ & $1$ & $j$ \\\hline
    $[010]$ & $2$ & $2$ & $3$ & $1$ & $j$ \\\hline
    $[011]$ & $3$ & $1$ & $4$ & $1$ & $j$ \\\hline
    $[100]$ & $4$ & $2$ & $4$ & $1$ & $j$ \\\hline
    $[101]$ & $5$ & $3$ & $4$ & $1$ & $j$ \\\hline
    $[110]$ & $6$ & $1$ & $2$ & $-1$ & $-j$ \\\hline
    $[111]$ & $7$ & $1$ & $3$ & $-1$ & $-j$ \\\hline
    \end{tabular}
    \vspace*{-0.4cm}
    \label{ciz1}
\end{table}

After the formation of $\mathbf{s}$ as above, it is then spread by multiplication with a predefined codebook matrix $\mathbf{C}\in\mathbb{C}^{N\times M}$ whose elements are randomly generated and Bernoulli distributed to enable CS-based detection at the receiver \cite{ji2018sparse}. After spreading $\mathbf{s}$, the transmitted OFDM block $\mathbf{x}_F\in\mathbb{C}^{N\times 1}$, whose all subcarriers are non-zero, is obtained as:
\begin{equation}
\mathbf{x}_F=\begin{bmatrix} x(1) & x(2) & \cdots & x(N) \end{bmatrix}^\mathrm{T}=\mathbf{Cs}.
\label{Eq.1}
\end{equation}

After this point, the same procedures for OFDM are applied. The inverse fast fourier transform (IFFT) is applied to the OFDM block to obtain to the time domain OFDM block $\mathbf{x}_T$:
\begin{equation}
    \mathbf{x}_T=\mathrm{IFFT}\big\{\mathbf{x}_F\big\} = \begin{bmatrix} X(1) & X(2) & \cdots & X(N) \end{bmatrix}^\mathrm{T}.
\label{Eq.2}    
\end{equation}
\noindent
After the IFFT operation, a cyclic prefix (CP) of length $L$ samples $\begin{bmatrix} X(N-L+1) & \cdots & X(N-1)X(N) \end{bmatrix}^\mathrm{T}$ is added to the beginning of the OFDM block. Once parallel to serial (P/S) and digital/analog conversions are applied, the signal is transmitted over a frequency-selective Rayleigh fading channel, which can be represented by the channel impulse response (CIR) coefficients
\begin{equation}
\mathbf{h}_T=\begin{bmatrix} h_T(1) & \cdots & h_T(v) \end{bmatrix}^\mathrm{T}\textcolor{red}{,}
\label{Eq.3}
\end{equation}
where $h_T(\alpha),\alpha=1,\dots,v$ are circularly symmetric complex Gaussian random variables with the $\mathcal{C}\mathcal{N}(0,\frac{1}{v})$ distribution. With the assumption that the channel remains constant during transmission of an OFDM block and the CP length $L$ is larger than $v$, the equivalent frequency domain input-output relationship of this OFDM scheme is given by 
\begin{equation}
    y_F(\beta)=x_F(\beta)h_F(\beta)+w_F(\beta), \hspace{0.2cm}       \beta=1,\dots,N
\label{Eq.4}    
\end{equation}
where $y_F(\beta)$, $h_F(\beta)$ and $w_F(\beta)$ are the received signals, the channel fading coefficients and the noise samples in the frequency domain, whose vector forms are given as $\mathbf{y}_F$, $\mathbf{h}_F$ and $\mathbf{w}_F$, respectively. The distributions of $h_F(\beta)$ and $w_F(\beta)$ are $\mathcal{C}\mathcal{N}(0,1)$ and $\mathcal{C}\mathcal{N}(0,N_{0})$, respectively, where $N_{0}$ is the noise variance in the frequency domain, which is equal to the noise variance in the time domain. The signal-to-noise ratio (SNR) is defined as $E_b/N_{0}$, where $E_b=(N+L)/ m$ is the average transmitted energy per bit. The spectral efficiency of the proposed scheme is
\begin{equation}
    \eta =m/(N+L) \hspace{0.2cm} \text{[bits/s/Hz]}.  \vspace{-0.4cm}
\label{Eq.5}    
\end{equation}\

\subsection{E-SVC-OFDM Decoder} 
At the receiver side, after removing CP and performing FFT, the MMP-DF algorithm \cite{kwon2014multipath}, whose inputs are given in Table II, is performed to decode $\mathbf{y}_F$ under the assumption of perfect channel state information. MMP-DF is chosen as the algorithm used due to its precise sparse vector recovery and controllable complexity based on the number of candidates desired for approximation. The MMP-DF algorithm outputs a sparse vector  $\mathbf{\hat{s}}\in\mathbb{C}^{M\times 1}$ and works by using a tree search with the help of a greedy search method. As long as the performance gain is high at low tree sizes, then the usage of MMP is well\textcolor{red}{-}justified \cite{kwon2014multipath}. $K$ is fixed to two for efficient operation of the algorithm as in \cite{ji2018sparse}. To be able to apply the MMP-DF algorithm, $\mathbf{y}_F$ and $\mathbf{h}_F$ are modified and used as an input\textcolor{red}{s} for the algorithm. The modified received vector can be expressed
as\cite{ji2018sparse}
\begin{equation}
    \hat{\mathbf{y}}_F=\mathrm{diag}(\hat{\mathbf{h}}_F) \mathbf{y}_F\textcolor{red}{,}
\label{Eq.6}    
\end{equation}
where $\mathrm{diag}(\mathbf{\cdot})$ creates a diagonal matrix from a vector, $\hat{\mathbf{h}}_F=\begin{bmatrix}   e^{j\angle{h_F(1)}} & \cdots & e^{j\angle{h_F(N)}} \end{bmatrix} ^\mathrm{T}$ and $\angle{h_F(\beta)}$ is the angle of $h_F(\beta)$. The sensing matrix can be expressed as
\begin{equation}
    {\boldsymbol{\Psi}} = \mathrm{diag}(\mathbf{\hat{h}}_F\odot\mathbf{h}_F) \mathbf{C}\textcolor{red}{,}
\label{Eq.7}    
\end{equation}

\begin{table}[t]
      \renewcommand{\arraystretch}{1.1}%
      \centering
      \caption{MMP-DF Inputs and Parameters}
      \begin{tabular}{|>{\centering\arraybackslash}m{4cm} | >{\centering\arraybackslash}m{2cm}|} 
      	
      \hline
      Inputs & Notation \\ [0.5ex] 
      		
      \hline\hline
      Modified received vector & $\mathbf{\hat{y}}_F$  \\\hline
      Sensing matrix & ${\boldsymbol{\Psi}}$  \\\hline
      Number of active subcarriers & $K$  \\\hline
      Number of search expansions & $\Omega$  \\\hline
      Stop threshold & $\Lambda$  \\\hline
      Maximum iterations & $\Upsilon$  \\\hline
      \end{tabular}
      \label{ciz1}
\end{table} 
\noindent where $\odot$ stands for element-wise multiplication. Other inputs are the numbers of active subcarriers $K$, number of search expansions $\Omega$, stop threshold value $\lambda$ and maximum iterations $\Upsilon$. With all these inputs, MMP-DF algorithm is run and outputs $\hat{\mathbf{s}}$ that only includes $K$ non-zero elements \cite{kwon2014multipath}.
 
The AP indices $(\hat{i}_1, \hat{i}_2, \cdots, \hat{i}_K)$, which are obtained from $\mathbf{\hat{s}}$, are applied to the combinatorial algorithm \cite{bacsar2013orthogonal} to obtain the decimal number $\hat{d}$. If $2^{{ \lfloor a\rfloor}+1}-\binom{M}{K} \leq \hat{d}$, meaning $(\hat{i}_1, \hat{i}_2, \cdots, \hat{i}_K)$ are non-reused indices, $\hat{d}$ can be directly converted to bits. Otherwise, the two vectors $\mathbf{b}_1$ and $\mathbf{b}_2$ that contain the symbols $(x_1, x_2, \cdots, x_K)$ and the extended ones $(\Tilde{x}_1, \Tilde{x}_2, \cdots, \Tilde{x}_K)$, respectively, are used for detection. For this purpose, the vector $\mathbf{\hat{b}}=\begin{bmatrix}  \hat{b}_1 & \hat{b}_2 & \cdots & \hat{b}_K \end{bmatrix}^\mathrm{T}$, which includes the non-zero elements of $\mathbf{\hat{s}}$, is constructed. Finally, in the following, two decision metrics are calculated to find whether the original symbols or extended ones are employed:
\begin{equation}
    \hat{l}=\arg\underset{l \in {1,2}}{\min}\| \mathbf{\hat{b}}-\mathbf{b}_l\|^2.
\label{Eq.8}    
\end{equation}

If $\hat{l}=1$, corresponding to the original symbols, then $\hat{d}$ is directly converted to bits. On the other hand, if $\hat{l}=2$, which represents the extended symbols,  $\hat{d}+\binom{M}{K}$ is converted to bits. To further illustrate the E-SVC-OFDM decoder, a simple example is shown in Fig. 3.

\begin{figure}[t]
		\centering
		\includegraphics[width=1\columnwidth,height=5.7cm]{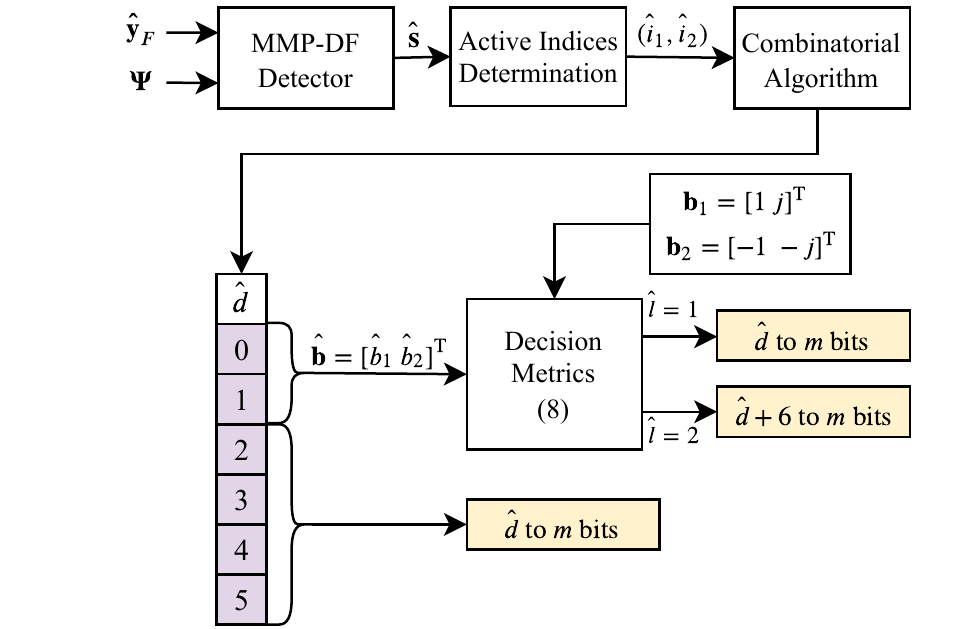} 
		\caption{A flow diagram example of the E-SVC-OFDM decoder for $M=4$ and $K=2$.} 
		\label{Fig. 4}
\end{figure}

\section{Codebook Index Modulation} 
In this section, we propose the novel scheme of SE-CBIM whose block diagram is shown in Fig. 2. SE-CBIM is a waveform that further improves the spectral efficiency of the reliable and low-latency E-SVC-OFDM system. This method conveys information bits by not only the indices of active VDD components but also the selected codebook index.
\subsection{SE-CBIM Encoder}
 The transmitter of the proposed system is built on to that of E-SVC-OFDM, with a major difference due to the use of multiple codebooks instead of one. As shown in Fig. 2, the incoming $m$ bits are split into $m_1=\log_2{G}$ and $m_2={ \lfloor a\rfloor}+1$ bits. Here, $m_1$ bits choose the $g$th codebook $\mathbf{C}_g$, $g\in \{1,\dots,G\}$, where $G$ is the total number of predefined codebook matrices whose elements are randomly generated with Bernoulli distribution and is an integer power of two. Then, $m_2={ \lfloor a\rfloor}+1$ bits are applied to the E-SVC-OFDM encoder as previously done to determine $\mathbf{s}$. Therefore, in SE-CBIM, for $G$ codebooks, a total of 
 \begin{equation}
   m= \log_2{G}+{\lfloor a\rfloor}+1\textcolor{red}{,}
\label{Eq.9}   
\end{equation}
bits can be transmitted for each OFDM symbol.


\subsection{SE-CBIM Decoder}
At the receiver side, the MMP-DF algorithm is run $G$ times for each generated codebook resulting in $G$ number of output vectors $\mathbf{\hat{s}}_g$ with the set of active indices $\Bar{I}_g=(\hat{i}_{g,1}, \hat{i}_{g,2}, \cdots, \hat{i}_{g,K})$, where $\hat{i}_{g,k}$ is the $k$th active index of the output vector corresponding to the $g$th codebook. For each $\mathbf{\hat{s}}_g$, two vectors $\bar{\mathbf{b}}_{g,1}$, $\bar{\mathbf{b}}_{g,2}\in\mathbb{C}^{M\times 1}$, which include the elements of $\mathbf{b}_1$ and $\mathbf{b}_2$, respectively, are produced using the indices in $\Bar{I}_g$.
 Then, the decoded codebook index $\hat{g}$ is determined by calculating a total of $2G$ metrics as follows:
\begin{equation}
    (\hat{g},\hat{l})=\arg\underset{g,l\in {1,2}}{\min} 
    \| \mathbf{\hat{s}}_{g} - \bar{\mathbf{b}}_{g,l} \|^2.
\label{Eq.10}    
\end{equation}
According to $\hat{g}$, $m_1$ bits are decoded directly. Then, the active indices $\Bar{I}_{\hat{g}}$ of the determined codebook are applied to the combinatorial algorithm to obtain the decimal value $\hat{d}$. If $2^{{ \lfloor a\rfloor}+1}-\binom{M}{K} \leq \hat{d}$ or if $\hat{d}<2^{{ \lfloor a\rfloor}+1}-\binom{M}{K}$ and $\hat{l}=1$, $\hat{d}$ is directly converted to $m_2$ bits. Otherwise, if $\hat{d}<2^{{ \lfloor a\rfloor}+1}-\binom{M}{K}$ and $\hat{l}=2$, meaning the symbols are extended, $\hat{d}+\binom{M}{K}$ is converted to $m_2$ bits.

\section{Complexity Analysis}
\begin{figure}[t]
		\centering
		\includegraphics[width=1\columnwidth,height=6cm]{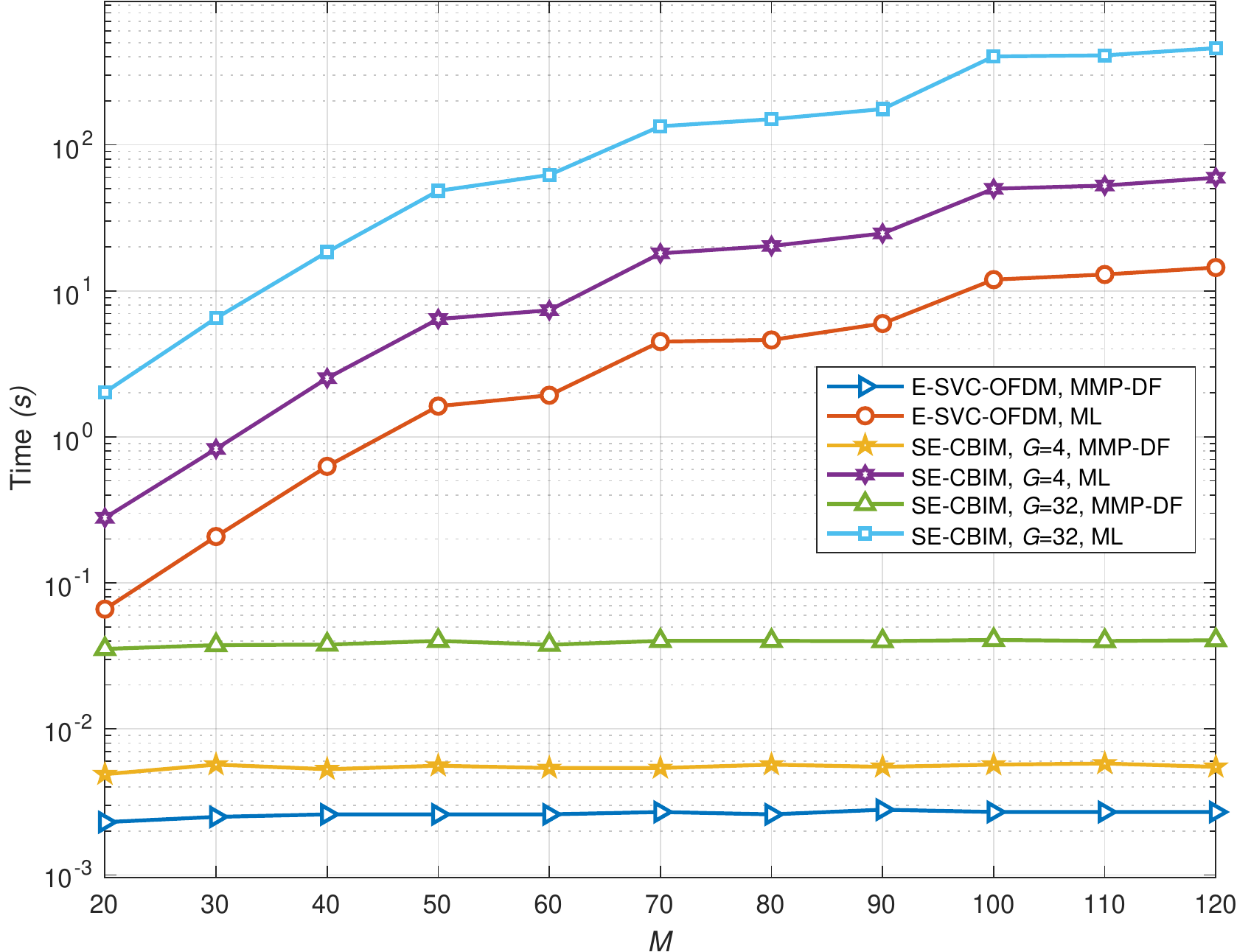} 
		\caption{Running time as a function of $M$ where the proposed schemes use ML and MMP-DF detection.}
		\vspace*{-0.4cm}
		\label{Fig. 5}
\end{figure}
In this section, we analyze the complexity of  E-SVC-OFDM and SE-CBIM detectors in terms of the running time as in \cite{kwon2014multipath}. We have exploited the MATLAB software on a computer with an Intel Core i7-3930K processor and Microsoft Windows 10 environment to obtain the running time. In \cite{kwon2014multipath}, it is stated that MMP-DF algorithm is comparable with other greedy algorithms such as StOMP \cite{StOMP} and CoSaMP \cite{COSAMP} when $K$ is low. In our proposed schemes, we have selected $K=2$, therefore, the algorithm we employ can be considered as a low-complexity one. Also, since channel coding is not required to achieve outstanding BER in SE-CBIM, it is relatively a much less complex scheme when compared to channel coded ones. To give further insight into the decoding complexity, we show the comparison of ML detection and the proposed detectors in Fig. 4. Assume that the set including all possible sparse vectors in VDD is given by $\mathcal{X}=\{\mathbf{x}_1,\cdots,\mathbf{x}_{2^{\lfloor a \rfloor+1}}\}$. Then, by employing $\mathcal{X}$, ML detection rules for E-SVC-OFDM and SE-CBIM can be obtained, respectively, as:
\begin{equation}
    \hat{\mathbf{x}}=\arg\underset{\mathbf{x} \in \mathcal{X}}{\min} 
    \| \mathbf{y}_F - \mathbf{h}_F \odot (\mathbf{C}\mathbf{x}) \|^2.
\label{Eq.11}    
\end{equation}
\begin{equation}
    (\hat{\mathbf{x}},\hat{g})=\arg\underset{g,\mathbf{x} \in \mathcal{X}}{\min} 
    \| \mathbf{y}_F - \mathbf{h}_F \odot (\mathbf{C}_g\mathbf{x}) \|^2.
\label{Eq.12}        
\end{equation}
As seen from Fig. 4, ML detection is  inefficient in terms of complexity for the proposed schemes. However, MMP-DF algorithm presents much lower complexity compared to ML method. In Fig. 4, since $K$ is low, the MMP-DF detector is more efficient than the ML detector as in \cite{kwon2014multipath}. Also, it can be observed that with an increase of the number of codebooks, the decoding complexity increases. It should be noted that the complexity of the decision metric calculations in (\ref{Eq.8}) and (\ref{Eq.10}) after the MMP-DF algorithm to determine whether the symbols are extended, is negligible and can be ignored. Additionally, in contrast to ML detection, MMP-DF is not effected by $M$ since $K$ is constant. 

\section{Simulation Results} 

In this section, computer simulation results are presented for varying parameters of E-SVC-OFDM and SE-CBIM. Additionally, a BER comparison is shown for convolutional coded OFDM \cite{LTE}, E-SVC-OFDM and SE-CBIM. Error performance of these schemes are obtained through Monte Carlo simulations. For all simulations, a $10$-tap frequency-selective Rayleigh fading channel and two active indices ($K=2$) in the VDD is assumed, where $x_1=1$, $x_2=j$ and $\Tilde{x}_1=-1$, $\Tilde{x}_2=-j$. The CP length is set as $16$ and the LTE convolutional code with rate $1/3$ was chosen \cite{LTE}. The parameters $\Omega$, $\Lambda$ and $\Upsilon$ for the MMP-DF algorithm are set as $2$, $0.1$, and $2$, respectively.
	
The first simulation involves only the E-SVC-OFDM scheme with a varying number of subcarriers $(N)$ in the frequency domain as $M=128$ is kept constant, which changes the spectral efficiency. As seen from Fig. 5(a), as the number of elements in the virtual domain is kept constant and the number of subcarriers are escalated from $32$ to $512$, a noticeable improvement is observed in BER performance due to the APs of the sparse vector being spread to more resources in the frequency domain. In other words, spreading the APs to more subcarriers provides a coding gain. As $N$ increases, the correlation between randomly generated codewords decreases, and this helps with better recovery of sparse vectors. However, spectral efficiency decreases as the number of subcarriers increases because more spectrum is used with more subcarriers. It is also seen in Fig. 5(a) that the BER improvement reaches a saturation while the largest improvement occurs until $N\leq M$. 
	    
In Fig. 5(b), the number of subcarriers is fixed to $128$, and the number of elements in the virtual domain ($M$) is varied as $32, 64$ and $128$. It is observed that with increasing the size of the virtual domain, the error performance improves relatively. This can be explained by the fact that as $M$ increases, sparsity increases resulting in an improvement in BER and an increase in the number of transmitted bits. 

\begin{figure}[t]
		\centering
		\includegraphics[width=0.95\columnwidth,height=6cm]{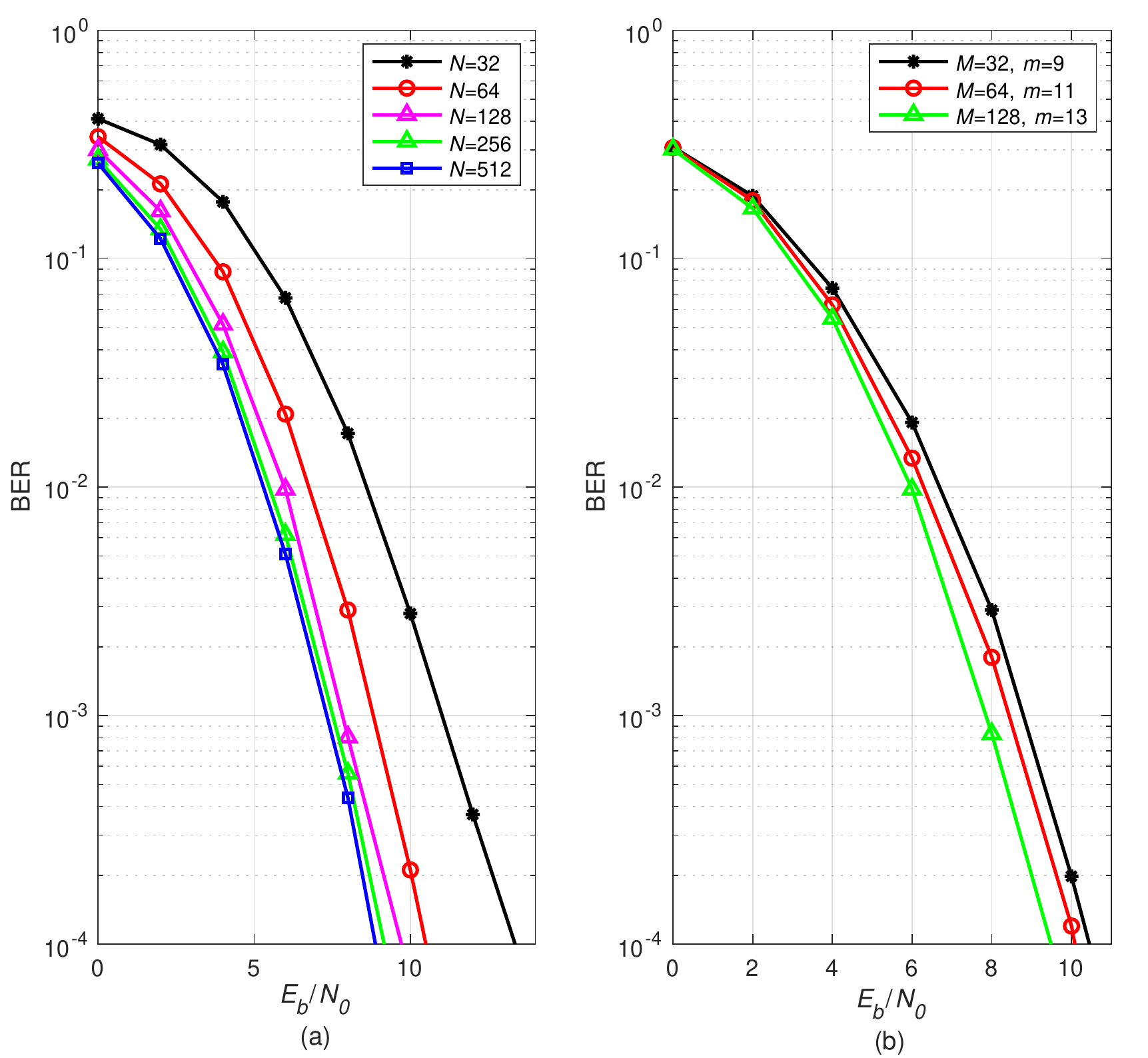}
		\vspace*{-0.2cm}
		\caption{Error performance of E-SVC-OFDM (a) for $M=128$ and varying $N$ and (b) for $N=128$ and varying $M$.}
		\vspace{-0.3cm}
		\label{Fig. 3}
\end{figure}  
\begin{figure}[t]
		\centering
		\includegraphics[width=0.95\columnwidth,height=5.95cm]{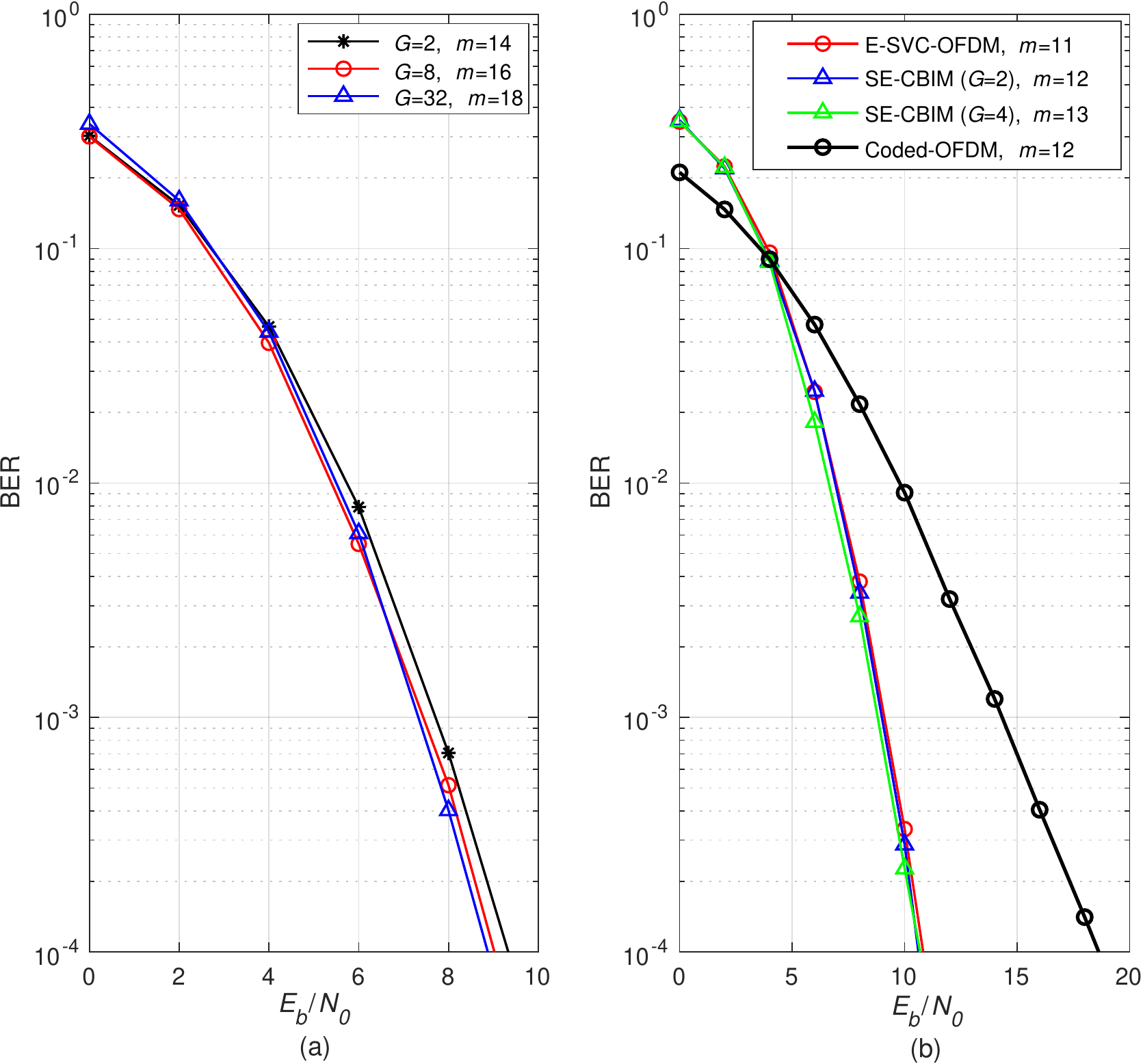} 
		\vspace*{-0.2cm}
		\caption{Error performance of SE-CBIM (a) for $N=128, M=128$ and (b) the comparison of relevant schemes for $N=64$ and $M=64$.}
		\vspace{-0.3cm}
		\label{Fig. 4}
\end{figure}

Fig. 6(a) presents the improvements in BER as the number of codebooks are increased in the proposed SE-CBIM scheme. It is shown that with this increase, the number of bits transmitted increases and this also provides a slightly better BER due to the characteristics of IM. However, there exists a trade-off: as the number of codebooks and bits transmitted is increased, the latency also increases due to additional codebook possibilities searched at the receiver as discussed in Section IV. It is also worth noting that the spectral efficiency values in the simulations are in the level of plain SVC-OFDM\cite{ji2018sparse}. 

Finally, our scheme is compared to the existing LTE convolutional coded OFDM \cite{LTE} with zero padding in Fig. 6(b). We have used zero padding for coded OFDM to obtain the same spectral efficiency with the proposed scheme and employed random interleaving to improve its performance. SE-CBIM provides an outstanding BER with an improvement of $10$ dB compared to coded-OFDM and relatively lower complexity due to the requirement of coding being removed. Additionally, it is seen that with SE-CBIM, without reducing the error performance, additional bits can be transmitted by exploiting codebook indices with increased complexity. Consequently, SE-CBIM presents an interesting trade-off between complexity and the number of bits transmitted by IM.

\section{Conclusion}		
In this paper,  E-SVC-OFDM and SE-CBIM schemes have been presented as reliable and low-latency waveforms. E-SVC-OFDM provides a clever and efficient use of the APs compared to the existing SVC-OFDM. SE-CBIM has been introduced as a novel waveform with increased spectral efficiency by utilizing IM compared to conventional SVC-OFDM schemes. The improved BER of the SE-CBIM scheme over the existing coded-OFDM scheme has been demonstrated through Monte Carlo simulations. Further reliability and latency improvements with generalized designs may be possible and left as future work.
	
	\vspace{-0.2cm}
	\balance
\bibliographystyle{IEEEtran}
\bibliography{IEEEabrv,references}

\end{document}